\documentclass{article}

\usepackage{microtype}
\usepackage{graphicx}
\usepackage{subfigure}
\usepackage{booktabs} 

\usepackage{algorithm}
\usepackage{algpseudocode}

\usepackage{epsfig}
\usepackage[accepted,nohyperref]{icml2024}

\usepackage{amsmath}
\usepackage{amssymb}
\usepackage{mathtools}
\usepackage{amsthm}

\usepackage[capitalize,noabbrev]{cleveref}

\theoremstyle{plain}

\theoremstyle{definition}

\theoremstyle{remark}

\usepackage[textsize=tiny]{todonotes}
\usepackage[accepted]{icml2024}[accepted]

\icmltitlerunning{Correctness Verification of Neural Networks Approximating Differential Equations}

\begin{document}

\twocolumn[
\icmltitle{Correctness Verification of Neural Networks Approximating Differential Equations}



\begin{icmlauthorlist}
\icmlauthor{Petros Ellinas}{yyy}
\icmlauthor{Rahul Nellikath}{yyy}
\icmlauthor{Ignasi Ventura}{yyy}
\icmlauthor{Jochen Stiasny}{yyy}
\icmlauthor{Spyros Chatzivasileiadis}{yyy}
\end{icmlauthorlist}

\icmlaffiliation{yyy}{Department of Wind and Energy Systems, 
 Technical University of Denmark (DTU)\\
 Elektrovej, 2800 Kgs. Lyngby, Denmark\\}

\icmlcorrespondingauthor{Petros Ellinas}{petrel@dtu.dk}

\icmlkeywords{Correctness Verification, Partial Differential Equations, Complete Verification, Gradient Attack, Bound Propagation}

\vskip 0.3in
]




\begin{abstract}
Verification of Neural Networks (NNs) that approximate the solution of Partial Differential Equations (PDEs) is a major milestone towards enhancing their trustworthiness and accelerating their deployment, especially for safety-critical systems. If successful, such NNs can become integral parts of simulation software tools which can accelerate the simulation of complex dynamic systems more than 100 times. However, the verification of these functions poses major challenges: it is not straightforward how to efficiently bound them or how to represent the derivative of the NN. This work addresses both these problems. First, we define the NN derivative as a finite difference approximation. Then, we formulate the PDE residual bounding problem alongside the Initial Value Problem's error propagation. Finally, for the first time we tackle the problem of bounding an NN function without a-priori knowledge of the output domain. For this, we build a parallel branching algorithm, that combines the incomplete CROWN solver and Gradient Attack for termination and domain rejection conditions. We demonstrate the strengths and weaknesses of the proposed framework, and we suggest further work to enhance its efficiency.
\end{abstract}

\section{Introduction}
Neural Networks (NNs) have been widely used as function approximators in several domains, such as forecasting \cite{forecasting1,forecasting2}, optimization \cite{nellikkath2022minimizing,NELLIKKATH2022108412} or control \cite{control1}. The benefits of using NNs are many and vary from their ability to adapt their representations to unseen data to their potential to be fast and highly accurate non-linear parametric approximations of known and unknown functions, if well trained. While this allows us to approximate a wide range of complex functions, general statements on the accuracy of the learned approximation are difficult to obtain. 

Correctness guarantees or correctness verification aim at providing a formal bound on the lowest accuracy across the relevant input domain. The concept behind correctness guarantees involves determining the worst-case approximation error in the input domain $\mathcal{D}$ and it can be formulated as an optimization problem
\begin{equation} \label{unique_form}
    \max_{x \in \mathcal{D}} | u(x) - u_{\theta}(x) |,
\end{equation}
where $u(x)$ is the ground truth solution, and $u_{\theta}(x)$ is the NN function approximation with weights $\theta$. Here, $x \in \mathcal{D}$ is a point in the input domain $\mathcal{D}$. The argument that maximizes \eqref{unique_form} indicates where the approximator has the worst performance compared to the ground truth. To enable the application of NNs in safety critical applications such as power systems, providing such guarantees is essential. 


A NNs application with remarkable potential involves approximating solutions $u_\theta(x)$ for Partial Differential Equations (PDEs), moving away from the strong reliance on classical numerical methods. PDEs provide a very powerful modeling approach for explaining complex phenomena in science and engineering. Therefore, a crucial subset of the correctness verification problems involve finding the worst-case approximation error, when approximating the solution to PDEs with NNs. Consequently, effective and scalable verification methods will play a key role in the ongoing advancements of deep learning methodologies, theories, and algorithms \cite{blechschmidt2021ways}. 

One of the most significant training frameworks for approximating general PDEs, namely Physics Informed Neural Networks (PINNs), was proposed in \citet{RAISSI2019686}. Here, the authors included the underlying physical equations in the NN training. NNs trained in this manner have seen numerous applications in the literature to approximate steady and unsteady PDEs \cite{cuomo2022scientific}.
PINNs also have been applied to the solution of Ordinary Differential Equations (ODEs) for engineering applications \cite{stiasny2023physics, stiasny2023pinnsim,Nath2023} in order to speed up time-domain simulations by at least 10-100 times. Furthermore, in \cite{Nath2023}, a PINN was suggested to monitor diesel engine health, assess engine dynamics, and predict maintenance needs, predicting the ODE system solution.

The verification of such NN-based PDE approximators faces a number of limitations, namely their computational efficiency and the reliance on differentiable activation functions in the NN \cite{eiras2023provably}. Furthermore, as the ground truth solution $u(x)$ is generally not attainable over the entire input domain, we revert to verifying necessary conditions, defined by $g(u_{\theta}(x))$ \cite{eiras2023provably}
\begin{equation} \label{corr_ver}
    \max_{x \in \mathcal{D}} | g(u_{\theta}(x)) |
\end{equation}
Since these functions or operators can involve derivatives of $u_{\theta}(x)$, the verification complicated.

Complete verification frameworks proposed in the literature such as $\alpha \beta$-CROWN \cite{generalbab}, aim on finding the solution to problems such as \eqref{corr_ver}, for a small input domain $\mathcal{D}$ and for some predefined points $x_0$. To this end, incomplete verifiers as $\alpha$-CROWN \cite{xu2021fastalphacrown} are combined with a Branch and Bound method. However, in these problems the domain of $g(u_{\theta}(x))$ is known beforehand. Therefore, the verification condition of the domains alongside the termination condition of the algorithm are provided a-priori. However, in the case of verifying the NNs as PDE approximators, the output domain is not foretold. Therefore there is the need of defining the conditions that will allow us to discard some domains as verified, while having their bounds.

Moreover, the large memory requirements of the NN training procedure have led to the creation of a new generation of energy and memory-efficient NN architectures with diverse activation functions \cite{bai2019deep} and new quantization techniques. However, these novel architectures occasionally require the transformation of fully differentiable activation functions to non-fully differentiable activation functions \cite{zhang2023artificial}. The presence of such activation functions in NNs has further complicated the NN correctness verification framework. Consequently, it is desirable for verification methods to support general activation functions.

In this work, we propose adaptations to the existing verification algorithms that can be used for correctness verification problems involving differential operators and non-differentiable activation functions. We achieve this by utilizing a Finite Difference scheme for the approximation of involved derivatives and furthermore apply a branching technique that improves the bound tightness. Equally importantly, we provide novel termination and node rejection guarantees for the Branch and Bound algorithm, using Gradient attack techniques. We demonstrate the algorithm on two PDEs. Additionally, we test the algorithm for an ODE solution, as the obtained bounds from \eqref{corr_ver} can then be used to bound the approximation $| u(x) - u_{\theta}(x) |$.

\section{Related work}
\subsection{Robustness Verification}
Robustness Verification of NNs is a field that has been widely studied in domains such as Computer Vision \cite{mu2019mnist} and Natural Language Processing \cite{Shi2020Robustness}. The objective is to find changes in specific input samples that result in significant alterations to the output. This domain is classified into two categories: Complete Verification and Incomplete Verification. The result of incomplete verification is a loose lower bound to problem \eqref{corr_ver}, where $\mathcal{D}$ is considered a region around a specific input sample. On the other hand, complete verification provides the exact solution to problem.

Bound Propagation has been extensively studied \cite{gowal2019effectiveness,zhang2018efficient} in the context of incomplete verification of NNs, as it provides the bounds of the output neurons given an input domain. The incompleteness of these methods stems from the outer-approximation of the feasible domain of the non-linear activation functions. This approximation is used to reduce their computational burden. Interval Bound Propagation \cite{gowal2019effectiveness} and CROWN \cite{zhang2018efficient} were some of the methods that were proposed for this problem. These methods are building blocks of complete verifiers \cite{wang2021betacrown}, if they are combined with global optimization methods such as Branch and Bound \cite{1960BAB}. 

\subsection{Correctness Verification}
In contrast with robustness verification which aims to verify the stability of predictions within local regions around labeled points, correctness verification offers a method to verify the accuracy across the entire designated input space. It aims to validate the network's accuracy for all target inputs or identify areas where the NN's predictions are inaccurate \cite{yang2019correctness}. 

PINNs correctness verification has been studied from different angles. In \citet{DeRyck2022,wang2022l2,deryck2023error}, the authors have analyzed the error of PINNs as approximators for PDEs with certain properties, giving possible solutions for \eqref{corr_ver}. However, the proposed solutions suffer from scalability and tightness issues \cite{eiras2023provably}. On the other hand, the work carried out in \citet{eiras2023provably} uses robustness verification tools for bounding certain properties of the PINN approximation that can be related to the approximation error. However, this framework, can only be used with fully differentiable activation functions, which limits its applicability. Moreover, the used branching algorithm branches through the whole input domain. Our proposed algorithm relates to \citet{eiras2023provably} but allows for general activation functions and improves the branching algorithm by discarding verified domains.

\section{Formulation}
In the following, we introduce the PDE problem formulation and the PINN training to obtain an NN-based solution approximation. As the verification problem in \eqref{unique_form} cannot be solved, we formulate necessary correctness conditions to assess the performance of NN-based PDE approximators.
\subsection{Partial Differential Equations}

A general partial differential equation (PDE) problem involves finding a function \(\boldsymbol{u}(x, t)\) that satisfies a PDE within a given spatial domain \(\Omega\) and a time domain [0,T] and is subject to the appropriate boundary and initial conditions. 

We consider the general form of a time-dependent PDE
\begin{equation}
    F\left(\boldsymbol{u}, \frac{\partial \boldsymbol{u}}{\partial t}, \frac{\partial \boldsymbol{u}}{\partial x}, \ldots, x, t\right) = 0, \text{for} \quad (x, t) \in \Omega \times [0,T]
\end{equation}
where \(F\) represents a differential operator combining various derivatives of \(u\) with respect to the spatial and the temporal variables. The problem is supplemented by a general form of Robin boundary conditions
\begin{equation}
    B[\boldsymbol{u}](x,t)=b_0(x,t) , \text{for} \quad (x, t) \in \partial \Omega \times [0,T]
\end{equation}
and initial conditions specifying the function's behavior at \(t = 0\)
\begin{equation}
    \boldsymbol{u}(x, 0) = u_0(x), \quad \text{for} \quad x \in \Omega.
\end{equation}

A subclass of PDE problems, that has engineering applications is the problem of Initial Value Problem. One main advantage of this special PDE category is that these problems have unique solutions, under some mild conditions.


\subsection{Physics Informed Neural Networks (PINNs)}
PINNs are a special framework of training NNs to approximate the solution of PDEs. The boundary and initial conditions alongside the PDE residual are evaluated on collocation points to form the loss function. Therefore we can denote the loss function of these NNs as:
\begin{equation}
    \mathcal{L}=\alpha \cdot \sum_{i \in N_r} F_i (\boldsymbol{u}_\theta)+ \beta \cdot \sum_{i \in N_b} B_i[\boldsymbol{u}_\theta] + \gamma \cdot \sum_{i \in N_u} (\boldsymbol{u}_\theta(x,0)-\boldsymbol{u}_0)
\end{equation}
where $\alpha,\beta,\gamma$ weight the different objectives in the loss function and $N_r, N_b, N_u$ are the sets of collocation points for the PDE residual, boundary, and initial condition. A NN $f_\theta$ with $n \in \mathbb N$ neurons and weights $\theta \in \mathbb{R}^n$, trained in this framework can be described as the approximation of the solution as
\begin{equation}
    f_\theta: (\boldsymbol{x_0}, t) \rightarrow \boldsymbol{u}_\theta
\end{equation}

\subsection{Essential Properties for Validating PDE Solution Approximations}

Correctness verification as formulated in \eqref{unique_form}, cannot be evaluated for PDE solution approximators as the ground truth solution $\boldsymbol{u}$ may not be unique, and it cannot be obtained analytically. However, \citet{eiras2023provably} formulate necessary properties of a PDE solution approximator to bound the approximation error
\begin{enumerate}
\item $\begin{aligned}[t]
    & \max_{x \in \Omega}|\boldsymbol{u}_{\theta}(x,0)-\boldsymbol{u}_0| \leq \delta
\end{aligned}$ \label{prop1}
\item $\begin{aligned}[t]
   & \max_{(x, t) \in \partial \Omega \times [0,T]}| B[\boldsymbol{u}_{\theta}](x,t)-b_0(x,t)| \leq \epsilon
\end{aligned}$ \label{prop2}
\item $\begin{aligned}[t]
    & \max_{(x, t) \in \Omega \times [0,T]}|F(\boldsymbol{u}_{\theta},\frac{\partial \boldsymbol{u}_{\theta}}{\partial t}, \frac{\partial \boldsymbol{u}_{\theta}}{\partial x}, \ldots,x,t)| \leq \zeta
\end{aligned}$  \label{prop3}
\end{enumerate}
As presented in \citet{eiras2023provably}, the obtained bounds $\delta$, $\epsilon$ and $\zeta$ can empirically be related to the approximation errors. We will discuss in Section 5 the special case of Initial Value Problems (IVPs), where the approximation error can be bounded using $\delta$ and $\zeta$.

\section{Correctness Verification of PDE approximators} \label{section_properties}

In this section, we first present  CROWN, which is a main ingredient for the proposed complete verification algorithm in this paper. Using this tool, one can calculate the bounds of the output of a function with respect to a continuous input domain. Second we propose arepresentation method for the NN derivatives, namely, the Finite Differences approximation. Finally, we introduce the input branching algorithm alongside the novel termination and node rejection conditions we propose in this paper. Our proposed algorithm improves tractability by applying two modifications, namely using finite difference approximations and input branching. Section 4.4 presents the resulting algorithm for complete correctness verification that this paper proposes..

\subsection{Bound Propagation with CROWN}
Bound propagation techniques seek to give an upper bound for the optimization problem \eqref{corr_ver}.  CROWN achieves that by backpropagating the bounds from the output to the input, with respect to the intermediate layers' bounds. Using CROWN for bound propagation is extremely effective because it is able to bound nonlinear activation functions with linear upper and lower bounds \cite{wang2021betacrown}. 


The additional property that makes the method attractive is the ability to efficiently solve the bound propagation problem using one or more GPUs, in contrast to a Linear Programming (LP) solver. Moreover, in opposition with the latter, CROWN can be used for a greater range of activation functions. Additionally it can be used in combination with other optimization methods, such as $\alpha$-CROWN \cite{xu2021fastalphacrown} or $\beta$-CROWN \cite{wang2021betacrown}, to achieve tighter bounds.

    
By bound propagation, one is trying to find a upper bound for the optimization problem 
\begin{equation} \label{bounds_prop}
    \max_{x \in C} f_\theta(x),
\end{equation}
where $f_\theta(x)$ expresses a NN with weights $\theta$. $C$ expresses an area around a point $\hat{x}$ as $C=\{\hat{x}: ||\hat{x}-x||_p\}$.
By forward and backward propagation, CROWN bounds each neuron's pre-activation value. Therefore, the algorithm finds a solution $\hat{f}$ to \eqref{bounds_prop} that is greater than the optimal solution. In our proposed approach, we will use bound propagation to find tighter bounds for properties \ref{prop1}-\ref{prop3} (see Section 3.3. 

To obtain the exact solution for the optimization \eqref{bounds_prop} one needs to use a Branch-and-Bound framework. Additionally, in \eqref{corr_ver}, functions $g$, may include the partial derivatives of the NN output. Concretely, in Properties \ref{prop2}, \ref{prop3}, the derivatives cannot be expressed trivially with computational graphs. Therefore, we propose to estimate these derivatives with the widely used Finite Difference Approximation, described below.

\subsection{Finite Difference Approximation with NNs}
\label{Sec:FDA}
Finite difference approximation is a numerical technique used to estimate derivatives or solve differential equations by approximating them with a discrete set of points. It involves dividing a continuous domain into a grid and replacing the derivatives in the equations with finite difference approximations derived from the function values at these grid points. By utilizing forward, backward, or central differences, depending on the specific application, this method transforms calculus problems into algebraic equations that can be solved computationally. Finite difference approximation finds extensive application in various fields like physics, engineering, finance, and computer science, providing a valuable tool for approximating solutions to differential equations that may lack analytical solutions or are too complex to solve directly. 

This finite differences method for calculating functions derivatives is widely used in the literature for solving IVPs and PDEs \cite{finitedifference}. As the task of bounding functions involves multiple partial derivatives of the parametric solution expressed by a NN, we use this method to find good approximations for the partial derivatives. The unique characteristic of this method, is that it requires a forward pass for
calculating the derivative. This makes it computationally expensive and unappealing for applications as calculating derivatives for back-propagation. On the other hand, it can be useful for applications such as bound propagation in function derivatives, due to the fact that CROWN makes use of both forward and backward propagation. The forward derivative of a parametric function $f_\theta$ can be expressed as
\begin{equation} \label{derivative}
    \frac{\partial f_\theta}{\partial x_i} = \lim_{h \rightarrow 0} \frac{f_\theta(x,x_i+h)-f_\theta(x, x_i)}{h}.
\end{equation}
Choosing a small enough $h$, one can approximate the derivative of a given NN. However, this poses limitations. As $h$ gets smaller, there can be overflow or underflow problems, causing round-off errors. A possible solution to this problem is to increase the number of bits representing numbers e.g. switching from 32-bit to 64-bit architecture. However, this will also increase the memory requirements.
\subsection{Input Branching}
An essential component of the proposed framework is input branching. As the provided bounds of CROWN are perturbation-dependent \cite{zhang2018efficient}, they can be too relaxed for a large input domain. Therefore a method is needed to reduce the perturbation used in CROWN. 

The input branching methodology, that was  proposed for this, falls into the Branch and Bound (BaB) framework \cite{1960BAB}, a widely recognized method for global optimization \cite{horst1996global} and NN verification \cite{jaeckle2021neural}. These algorithms are employed to guarantee the discovery of the global optima in the problem stated in \eqref{corr_ver}. This is achieved through iterative partitioning of the initial feasible set into smaller subsets (branching), while simultaneously calculating upper and lower bounds for the global maximum (bounding). 

Nevertheless, finding a valid stopping and node rejection condition for this application is not trivial, as we do not know the final bound a-priori. For this reason, in contrast to \citet{eiras2023provably}, we used a sample-less condition, that is based on Gradient Attack.


\subsubsection{Gradient Attack}
Finding the worst perturbation in a region can be formulated, as an unconstrained optimization problem in \eqref{bounds_prop}, and repeated here for convenience:
\begin{equation} \label{gradient_attack}
    \min_{x} f_\theta(x)  
\end{equation}
Many methodologies have been proposed to find a possible solution to \eqref{gradient_attack}, such as Fast Gradient Sign Method (FSGM) \cite{fastgradient} or Projected Gradient Descent (PGSM) \cite{madry2019deep}. In this work, we considered FSGM to find possible adversarial examples, but any method in the literature can be used. FSGM iterates as
\begin{equation}
    x_{t+1}=x_t-l \cdot \text{sign} ( \nabla f_{\theta} )
\end{equation}
where $x_{t}$ is the NN's input vector after $t$ iterations and $l$ is a constant called, learning step. The iteration begins from a starting point $x_0$, which usually affects the outcome. A major benefit of this methodology, is that it is massively parallelizable, for different starting points. This property can be used to increase the quality of the solution.

Using this methodology, we can acquire a local maximum of a function. Therefore, we can use this methodology, as a heuristic, to discard domains that have their maximum absolute upper bound value lower than the value obtained using gradient descent.





\subsection{Complete Correctness Verification}
Concerning the optimization expressed in \eqref{bounds_prop}, we can find its exact solution using a branch and bound algorithm. We propose a branching for the input domain of the function, shown in Algorithm~\ref{alg:cap}. Specifically, given a function $f_\theta(x)$, in each iteration the algorithm divides the input domain with bounds $[LB, UB]$ in $P$ equal pieces. Then, the algorithm runs two FSGM (one for each sign) to acquire the point with the worst performance in the area. Using CROWN, it calculates their upper and lower bounds, with respect to the bounds of the domain and discards the domains that have lower upper bound than the point given by FSGM. Each of the domains are stored in a heap, so the worst-bound domain can be acquired easily.

The algorithm continues until the first domain of the heap is a point that is found by the FSGM, or if the perturbation of this domain, has reached a predefined limit $\overline{\Theta}$. The perturbation of a domain can be defined as $\frac{(UB-LB)}{2}$. In Algorithm \ref{alg:cap}, $\kappa$ denotes the bound acquired by FSGM on 100 points which operates on a Region Defined by $[LB, UB]$. Moreover, $\beta_i$ are the bounds, acquired by CROWN, of the specific domains defined by $[LB_i, UB_i]$.

\begin{algorithm}[tb]
\caption{Input Branching Technique}\label{alg:cap}
\begin{algorithmic}
\Require $  f_\theta(x), LB, UB,\overline{\Theta}, P $ 

\State $\text{Get } \kappa \text{ solving \eqref{gradient_attack}}$
\State $Heap \gets (\kappa ,LB, UB)$
\State $Max \gets -\infty$
\State $\gamma \gets -\infty$
\While{$\frac{(UB-LB)}{2} > \overline{\Theta} \text{ or } \kappa= Max$} 
        \State $\text{Get } \kappa \text{ solving \eqref{gradient_attack}}$
        \If{$\kappa \geq \gamma$}
            \State $\gamma \gets \kappa$
        \EndIf
        \State $\text{Split } [LB,UB] \text{ into } P \text{ domains}$
        \State $\text{Run CROWN on the $P$ domains} \text{ and acquire } \beta_i$
        \For{$i \in P$}
            \If{$\gamma < \beta_i$}
                \State $Heap \cup (\beta_i ,LB_{i}, UB_{i})$
            \EndIf
        \EndFor   
        \State $\text{Get } (Max ,LB, UB) \text{ from Heap }$
    \EndWhile   
\end{algorithmic}
\end{algorithm}

\section{Using correctness verification for bounding the approximation error of IVPs}
A relevant subclass of PDEs are ODEs that occur in many engineering applications, usually as Initial Value Problems (IVPs). For this subset of problems, the previously derived properties can be used to bound the approximation error of learned solutions--the desired quantity we introduced in \eqref{unique_form}.  

An Initial Value Problem (IVP) is formulated using a set of Ordinary Differential Equations (ODEs) as:
\begin{align}
    F(\boldsymbol{u}(t),\frac{d\boldsymbol{u}(t)}{dt}, t) &= \frac{d}{dt}\boldsymbol{u}(t) - f(\boldsymbol{u}(t)) \label{first_general}\\
    u(0) &= u_0 \label{second_general}
\end{align}
where $\boldsymbol{u} \in \mathbb{R}^n$ denotes the n-dimensional vector of the system's state, and the time parameter $t$ lies within $[0,\tau)$. We assume that function $f: \mathbb{R}^n \rightarrow \mathbb{R}^n$ is a smooth differentiable function. ODEs are a special case of the general PDE formulation since the solution does not deviate through space; they are only time-dependent functions.


The solution $\boldsymbol{u}$ will be unique if we find the function $f$ to be globally or locally Lipschitz-continuous
\begin{equation} \label{lipsitz}
    ||f(a)-f(b)|| \leq C ||a-b||,
\end{equation}
where $C$ is the Lipschitz constant or function. If one finds the constant or function $C$ in \eqref{lipsitz}, the motion described by $f$ will be unique for a given initial state $u_0$. That is given by the Picard-Lindelöf theorem, which ensures the existence and uniqueness of the solution inside the time interval $t \in [0,\tau)$. The trajectory
\begin{equation} \label{solution_const}
    \boldsymbol{u}(t; \boldsymbol{u}_0)= u_0 + \int_{0}^{\tau} f(\boldsymbol{u}(t)) dt
\end{equation}
describes the solution of the IVP given its initial state $\boldsymbol{u}_0$.

Under these assumptions, we can then describe the approximation error, see \citet[Ch. I, Variant of Thm. 10.2]{Hairer}, for the IVP
\begin{equation} \label{time_error}
    e_{rr}(t) \leq e^{Ct} \delta +\int_{0}^{\tau}e^{C(\tau-t)} \zeta(t) dt.
\end{equation}
 The above has a significant value, as $\delta$ and $\zeta(t)$ correspond to bounding the functions in Properties \ref{prop1} and \ref{prop3}. This enables us to argue about the error dynamics of the NN-based approximator and can be a step to guarantee the quality of the approximation. 
\section{Numerical Results}
For the numerical evaluation of the proposed framework we have trained Physics-Informed Neural Networks to approximate the solution of two commonly used PDEs, namely, the Burgers equation and the Schrödinger equation. Additionally, we determined rigorous worst-case error guarantees for a PINN for an ODE synchronous generator model, which is widely used in Power Systems Time Domain Simulations. 

We compare the proposed branching algorithm with the state-of-the-art $\alpha \beta$-CROWN, for bounding Property \ref{prop3}, in all three benchmarks. We show that, using  $\alpha \beta$-CROWN's input splitting, $\alpha \beta$-CROWN can converge faster to the result in the PDE benchmarks. However, in the IVP benchmark, the proposed algorithm converges faster to the optimal value of the optimization. Morever, $\alpha \beta$-CROWN tries to verify a predefined condition on the output domain, e.g. it verifies that a neuron's value, in the output, is always less than 1, given the defined NN input domain. We show that this is not suitable for the applications that this paper targets. All the experiments took place in an Intel Xeon 24-core CPU with 72GB RAM for the proposed algorithm, and in a TeslaV100 GPU with 32GB memory, for running the $\alpha \beta$-CROWN.

\subsection{PDE Benchmarks}
In this subsection, we briefly present the PDE benchmarks we have used to evaluate our framework. Specifically, we briefly present the Burgers and Schrödinger equations.

\subsubsection{Burgers Equation}
Derived from the Navier-Stokes equations for velocity fields, this one-dimensional partial differential equation (PDE) is extensively used in mathematics, fluid dynamics, nonlinear acoustics, gas dynamics, and traffic flow. It operates within a temporal domain ranging from 0 to 1 and a spatial domain spanning from -1 to 1.

Burgers' PDE problem is expressed as:
\begin{equation} \label{Burgers_res}
    \frac{\partial u(x,t)}{\partial t} + u(x,t) \frac{\partial u(x,t)} {\partial x} - \frac{0.01}{\pi} \frac{\partial^2 u(x,t)}{\partial x^2}=0
\end{equation}
\begin{equation} \label{Burgers_init_conditions}
    u(0,x)+\sin{\pi x}=0
\end{equation}
\begin{equation} \label{Burgers_bound_conditions}
    u(t,-1)=u(t,1)=0
\end{equation}
\subsubsection{Schrödinger Equation}
The one-dimensional nonlinear Schrödinger equation (NLSE) is a fundamental equation in quantum mechanics and nonlinear optics, extensively studied for its rich mathematical properties and physical implications. It describes the behavior of wave packets in nonlinear media, where the wave function's evolution is influenced by both linear dispersion and nonlinear effects. Although it is 1-dimensional it is complex-valued, meaning that the approximator must approximate both real and imaginary parts. It is formulated as:
\begin{equation} \label{Shro_res}
    i \frac{\partial u(x,t)}{\partial t} + 0.5 \frac{\partial^2 u(x,t)}{\partial x^2}+ |u(t,x)|^2 u(t,x)=0
\end{equation}
\begin{equation} \label{Shro_init_conditions}
    u(0,x)-2sech(x)=0
\end{equation}
\begin{equation} \label{Shro_bound_conditions_1}
    u(t,-5)-u(t,5)=0
\end{equation}
\begin{equation} \label{Shro_bound_conditions_2}
    \frac{\partial u(-5,t)} {\partial x}-\frac{\partial u(5,t)} {\partial x}=0
\end{equation}

\subsection{Initial Value Problem Benchmark}
Next we present the swing equation for the Single Machine Infinite Bus system, a system widely used in power systems to simulate and study critical dynamic phenomena. The assumption is that the generator state does not directly influence the grid-side voltage. Therefore, a voltage reference is set in the point of the connection and the active power production $P$ is able to vary without affecting the reference. 
\begin{equation} \label{SMIB_FORMULATION}
    \left[
    \begin{matrix}
    \frac{d \delta(t)}{dt} \\
    \frac{d \Delta \omega}{dt}
    \end{matrix}
    \right] =
    \left[
    \begin{matrix}
    0 & 1 \\
    0 & -\frac{d}{m}
    \end{matrix}
    \right] 
    \left[
    \begin{matrix}
     \delta \\
    \Delta \omega
    \end{matrix}
    \right] +
    \left[
    \begin{matrix}
    0 \\
    \frac{1}{m} (P-V_1 V_2 B_{12} \sin(\delta))
    \end{matrix}
    \right]
\end{equation}
\begin{equation} \label{SMIB_init_cond}
    \delta(0) - \delta_0=0
\end{equation}
For this benchmark, we consider that $\Delta \omega_0 = 0.1$. The time domain is in the range $[0,2]$ and the $\delta_0$ takes values in $[0,1]$. Moreover, network parameters are set as considered to be $B_{12}=0.2$ p.u., voltages $V_1=1$ p.u. and $V_2=1$ p.u., machine's damping coefficient $d=0.15$ p.u., and machine inertia constant $m=0.4$ p.u. 

As this problem is an IVP with a specific Lipschitz constant, our goal is to find how the error evolves through time, as described in \eqref{time_error}. This is important as the ability, for the first time, to determine an ODE approximation error for the PINNs, would possibly establish PINNs as ODE solvers 
, leading to the creation of novel and faster time-domain simulators such as \citet{stiasny2023pinnsim}.

\subsection{Finite Difference Approximation Quality}
To monitor the derivatives' approximation quality across different values for $h$ in \eqref{derivative}, we sample $10^5$ points. We then evaluate the Mean Squared Error (MSE) of the partial derivatives, calculated by the proposed method and by automatic differentiation of the Burgers' equation NN approximators with ReLU and Tanh activation functions. 

\begin{figure*}[ht]
        \centering
            \subfigure[Approximation error with Tanh activation functions] 
            {
                \label{subfig:lab1}
                \includegraphics[width=.95\textwidth]{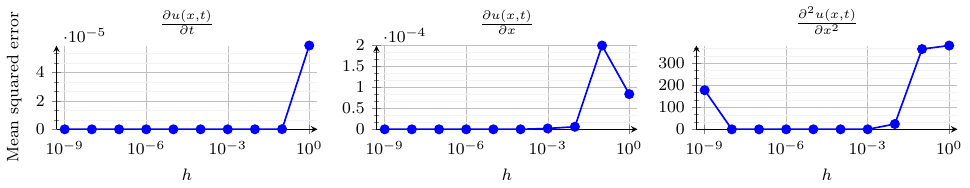} 
            } 
            \subfigure[Approximation error with ReLU activation functions] 
            {
                \label{subfig:lab4}
                \includegraphics[width=.95\textwidth]{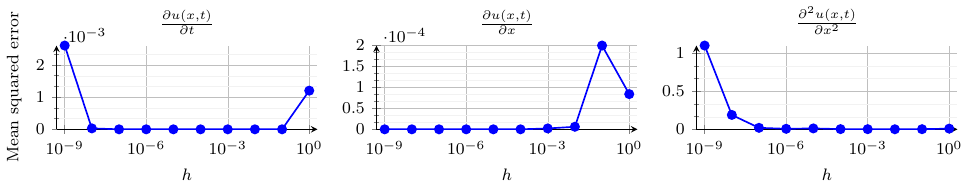} 
            }
        \caption{MSE of partial derivatives of NNs calulated by finite differences and automatic differentiation. Figure \ref{subfig:lab1} depicts the first order partial derivatives of time and the spatial variable alongside the second order spatial partial derivative for an NN approximator to the Burgers equation with Tanh activation functions. Correspondingly, in \ref{subfig:lab4} the partial derivatives of a NN approximator with ReLU activation functions are shown. }
        \label{fig:Figureders}
\end{figure*}

The results of this assessment can be seen in Figure \ref{fig:Figureders}. 
Specifically, from the assessment it can be seen that the difference between the two different methodologies for calculating the partial derivatives, is nearly zero, when h is taking values in $[10^{-2},10^{-7}]$. However, a major limitation of this methodology, as mentioned in \cref{Sec:FDA}, is the underflow or overflow problems, as the differences between the two terms in Finite Difference Approximation are small enough to cause it. This can be seen for small values of $h$, where the MSE increases. However, at the same time, $h$ must be chosen small enough to maintain sufficient quality of the derivative approximation. 

\subsection{Bounding Results}
In this section, we provide evidence indicating that the suggested algorithm achieves convergence to the actual solution at a faster rate compared to the leading tool, $\alpha \beta$-CROWN, for the IVP test case. We also highlight, that without the proposed termination and domain reduction conditions, $\alpha \beta$-CROWN is unsuitable for the correctness verification problems presented in this paper. The benchmarked $\alpha \beta$-CROWN cases, use either the Strong Branching (SB) heuristic with the input splitting property enabled or the nonlinear BaB heuristic \cite{generalbab} without input splitting. Due to extended computational duration, we have established a timeout of 18000s for verifying Property \ref{prop3} and 200s for Properties \ref{prop1} and \ref{prop2}. Moreover we benchmark the proposed methodologies with a FSGM from different starting points, which are taken from a grid sampling technique. A value of $l=10^{-3}$ was used for FSGM, we have set 100 points, spread in the whole input domain as starting points. Finally, we present the maximum value of the residuals, having as an input $10^{5}-10^{6}$ points, which acquired by grid sampling, the input domain. It is worth mentioning that for the $e(t)$ in Table \ref{sample-table-inco}, the numerical integration uses the Trapezoidal Rule, with 10000 points.

\begin{table*}[h]
\caption{Verified lower bounds for the $\alpha \beta$-CROWN Benchmarks within the set time limits.}
\label{sample-table-alpha}
\begin{center}
\resizebox{0.95\textwidth}{!}{%
\begin{small}
\begin{tabular}{lccccccc}
\toprule
Benchmark & Verified Conditions & $\alpha \beta$-CROWN-nonlinear & $\alpha \beta$-CROWN-SB & Proposed \\
\midrule
Burgers Equation &  \eqref{Burgers_res}  & $3.828 \times 10^9$ & $44.673$ & $2.808 \times 10^{6}$ \\ 
Schrödinger Equation & \eqref{Shro_res}  & $1.682 \times 10^8$ & $4.609$  & $3.617 \times 10^3$ & \\ 
Single Machine Infinite bus & \eqref{SMIB_FORMULATION} & $7.390 \times 10^{2}$ & $9.196 \times 10^{-2} $  & $3.091 \times 10^{-3}$ & \\
\bottomrule
\end{tabular}
\end{small}
}
\end{center}
\end{table*}
The proposed algorithm was built to support multiprocess parallelization. Therefore, it could harness the full power of provided CPUs. For the Property \ref{prop3}, the parallel process were 50 for each iteration. 
We use $h=10^{-6}$ for all finite difference approximations of first-order derivatives and $h = 10^{-3}$ for all second-order derivatives.
\begin{table*}[th]
\caption{Non-Verification Benchmarks}
\label{sample-table-inco}
\begin{center}
\resizebox{0.95\textwidth}{!}{%
\begin{small}
\begin{tabular}{lcccc}
\toprule
Benchmark & Verified Conditions & Sampling & Gradient Attack & Proposed \\
\midrule
Burgers Equation & $\eqref{Burgers_init_conditions}$ & $2.312 \times 10^{-3}$ & $1.460 \times 10^{-3}$ & $2.372 \times 10^{-3}$ \\
& $\eqref{Burgers_bound_conditions}$ & $1.932 \times 10^{-4}$ & $1.932 \times 10^{-4}$ & $2.046 \times 10^{-2}$ \\
& $\eqref{Burgers_res}$ & $9.996 \times 10^{-1}$ & $6.218 \times 10^{-2}$ & $2.808 \times 10^6$ \\
\midrule
Schrödinger Equation & $\eqref{Shro_init_conditions}$ & $2.230 \times 10^{-1}$ & $2.230 \times 10^{-1}$ & $2.230 \times 10^{-1}$ \\
& $\eqref{Shro_bound_conditions_1}$ & $3.178 \times 10^{-3}$ & $3.178 \times 10^{-3}$ & $3.178 \times 10^{-3}$ \\
& $\eqref{Shro_bound_conditions_2}$ & $2.420 \times 10^{-6}$ & $2.421 \times 10^{-6}$ & $2.591 \times 10^{-6}$ \\
& $\eqref{Shro_res}$ & $2.090 \times 10^0$ & $3.891 \times 10^{-1}$ & $3.617 \times 10^3$ \\
\midrule
Single Machine Infinite Bus & $\eqref{SMIB_init_cond}$ & $5.625 \times 10^{-1}$ & $5.625 \times 10^{-1}$ & $5.626 \times 10^{-1}$ \\
& $\eqref{SMIB_FORMULATION}$ & $3.086 \times 10^{-3}$ & $3.091 \times 10^{-3}$ & $3.091 \times 10^{-3}$ \\
& $e(t)$ & $8.430 \times 10^{-1}$ & $8.430 \times 10^{-1}$ & $8.430 \times 10^{-1}$ \\
\bottomrule
\end{tabular}
\end{small}
}
\end{center}
\end{table*}
In each process, CROWN was splitting each domain in 4 distinct parts, which implies that $P=2$. For the $\alpha \beta$-CROWN benchmarks, a single GPU was used, and the solver's batch size was set to 2048. All the other settings were left to their default value, except the Burgers equation case, in which we used the vanilla CROWN for all the verification process, instead of $\alpha$-CROWN. Furthermore, for the PDE benchmarks, we were seeking to verify the condition that the bound to the function representing Property \ref{prop3} is less than two. Similarly for the IVP Benchmark the condition was if the residual's bounds is less than $0.1$. In contrast, our proposed algorithm has a dynamic termination and domain discarding condition, which depend on the maximum value of the Gradient Attacks ran for every domain. Therefore, we do not require an approximation of the objective a-priori.

The following sections discuss the results of the Tables \ref{sample-table-alpha},\ref{sample-table-inco}. These tables present the lower bound that the shown algorithm reaches within the predefined time limit.

\subsubsection{ $\alpha \beta$-CROWN Benchmarks}
The comparison on convergence to the optimal value of the proposed methodology against $\alpha \beta$-CROWN is illustrated in Table \ref{sample-table-alpha}. As the most challenging tasks involved bounding Property \ref{prop3}, we have compared the convergence of the proposed algorithm against $\alpha \beta$-CROWN, only for Property \ref{prop3}. Their verification challenges stem from larger computational graphs and a broader input domain, causing incomplete verifiers to impose looser bounds. In the SMIB case, the proposed algorithm terminates in 281s and $\alpha \beta$-CROWN terminates in 453s. Nonetheless, in the PDE cases where both algorithms terminated due to the timeout limit, $\alpha \beta$-CROWN-SB, reaches a better bound. $\alpha \beta$-CROWN-nonlinear without input splitting, has the worst performance in all cases.
\subsubsection{Non-Verification Benchmarks}
In Table \ref{sample-table-inco}, we show the performance of the proposed algorithm against FSGM and Sampling Benchmark. Both benchmarked cases cannot be considered as verification. The reason is that with sampling techniques cannot give guarantees for the whole continuous domain, as they discretize it. However, as the test-cases' input domain is low-dimensional and small, we can trust that the Sampling benchmark would give a bound close the the actual bound, considering we perform a dense enough sampling. While FSGM does not guarantee the discovery of the global optima, it can be considered a good indicator for global optima in a low-dimensional input space. As detailed in the table, not every time the FSGM method is finding the worst-case error of the residuals. Consequently we choose to run it for every checked domain in Algorithm \ref{alg:cap}. Nonetheless, bounding the Properties \ref{prop1} and \ref{prop2} resulted in the algorithm terminating normally, having close results with Sampling, except of the bounding of \eqref{Burgers_bound_conditions}, which ended due to the time limit.
\section{Conclusions}
This paper targets the problem of correctness verification functions that include Neural Networks function approximators, especially when partial derivatives of NNs are involved. We formulated the correctness conditions for a general PDE approximator. We show that using these conditions we can bound the NN approximation error for an important subclass of problems, namely the Initial Value Problem, which has critical engineering applications. To bound these properties, we have shown that derivatives of NNs can be represented using finite differences approximation. Then we propose an input branching algorithm, with novel dynamic stopping and domain discard conditions, suitable for correctness verification problems, since their output domain is not know beforehand. Finally we empirically tested the methodology with multiple techniques

Bounding Property \ref{prop3} for the Burger's and Schrödinger's equation, leads to the creation of large computational graphs. This, in combination with the fact that the NNs have $\tanh{}$ as activation functions, leads to very loose bounds for CROWN, hence, the branching tree becomes very large. This prevents the node rejection with Gradient Descent solutions, leading to an overproduction of domains to be seen. Future work will focus on finding other heuristic-based node rejection techniques, such as proving the monotonicity of the domains, to restrict the number of branches. Ideas from $\alpha \beta$-CROWN can be used, to enhance the effectiveness of these conditions. 


\section{Impact Statement}
This paper presents work whose goal is to advance the field of Machine Learning. There are many potential societal consequences of our work, none which we feel must be specifically highlighted here.
\bibliography{icml2024/second_try}
\bibliographystyle{icml2024}

\newpage
\appendix

\onecolumn




\end{document}